\documentclass[12pt]{article}

\usepackage{amsmath,amssymb}
\usepackage[margin=1in]{geometry}
\usepackage{fancyhdr}
\usepackage{hyperref}
\usepackage{color}
\usepackage{booktabs, multirow, caption, subcaption}
\usepackage{setspace}
\usepackage{graphicx} 

\usepackage[backend=biber, style=apa, natbib=true]{biblatex}
\author{Akcan Balkir\thanks{Akcan Balkir, akcan.balkir@berkeley.edu, University of California, Department of Economics, 530 Evans Hall, Berkeley, CA 94720. I thank Alan Auerbach, Enrico Moretti, Mathilde Mu\~{n}oz, Emmanuel Saez, Joseph Shapiro, Reed Walker, and Danny Yagan for invaluable discussions and advice. I also thank Meredith Fowlie and numerous seminar participants at UC Berkeley. Funding from the Berkeley Opportunity Lab, the Fisher Center, and the Burch Center at UC Berkeley is thankfully acknowledged.}}
\title{Green Subsidies and Local Transitions:
Evidence from Energy Communities}
\date{Spring 2026}

\begin{document}

\maketitle

\begin{abstract}
This paper studies the effectiveness and incidence of the renewable energy Investment and Production Tax Credits. 
I leverage new geographical variation in these credits, introduced by the Inflation Reduction Act, to test whether renewable energy credits had real economic impacts. 
Communities with greater tax credits accrued 33\% more renewable energy capital and produced 31\% more renewable energy compared to similar counties. 
This suggests elasticities of 1.62 and 6.11 for the Investment and Production Tax Credits respectively.
I augment these results using an understudied dataset on planned investment to disentangle preplanned from additional projects.
Accounting for inframarginal investment significantly reduces the Investment Tax Credit's investment elasticity to 0.6.
After characterizing the supply side responses to these renewable tax credits, I document a new political feedback loop between increased incentives and support for renewable energy policies.
Areas with greater tax incentives experienced jumps in support for renewable energy policies, contrary to the Not In My Backyard narrative.
Heterogeneity in political responses suggests that the Investment and Production Tax Credits garnered support through two channels: 1) labor market spillovers, with construction wages increasing by 7\% in areas with greater tax incentives, and 2) public goods spillovers, with parents across party lines increasing support for renewable energy by 13\%.

\end{abstract}
\vfill
\pagebreak

\onehalfspacing

\section{Introduction}

Climate change is an urgent problem requiring immediate and large scale policy action to prevent irreversible damages to the environment and human welfare \citep{lee_ipcc_2023}. 
The textbook economic solution for climate change is a Pigouvian tax on carbon emissions \citep{pigou_economics_1932}, but carbon taxes face significant political opposition.
Instead, United States' climate policy has favored ``green'' subsidies over ``brown'' taxes \citep{borenstein_private_2012}.
For nearly half a century, the largest of these ``green'' subsidies have been the Investment and Production Tax Credits for renewable energy, despite unresolved questions about their impact and incidence.

Following the passage of the Inflation Reduction Act, estimates for the ten-year fiscal cost of the Investment and Production Tax Credits ranged from \$391 billion to \$1.2 trillion \citep{cbo_estimated_2022,della_vigna_carbonomics_2023}, highlighting the uncertainty regarding responses to these incentives. 
Traditional economic motivations for subsidizing renewable energy rely on positive externalities such as inducing innovation, learning by doing, and amplifying peer effects \citep{acemoglu_environment_2012, arrow_economic_1962,gerarden_role_2025}.
All of these motives assume that the Investment and Production Tax Credits create significant additional investments.
Yet there is no direct evidence on these credits' efficacy in generating real investment and production.
In fact, in the two years after the Inflation Reduction Act, U.S. renewable energy capital grew only 18.8\%, below the 19.7\% growth in the two preceding years.
Does this indict the centerpiece of America's green subsidies, or does it simply reflect changing macroeconomic conditions?
Disentangling these channels requires a method that can control for the business cycle.

A second set of questions concerns the physical differences in responses to infrastructure subsidies and Pigouvian taxation.
Unlike a carbon tax, which operates through prices, investment subsidies must be realized as physical infrastructure sited in specific communities, and a growing literature documents rising local opposition to renewable projects \citep{stokes_short_2020, klein_abundance_2025}. 
But support for renewable energy may be endogenous to investments: once projects are built the distribution of costs and benefits potentially reshape local attitudes toward the policy itself. 
Whether such feedback exists, and in which direction it runs, bears directly on the durability of second-best environmental policy and the long-run welfare comparison between subsidies and taxes.

This paper provides evidence on both sets of questions by studying the first ever geographical variation in the Investment and Production Tax Credits.
Since the Investment and Production Tax Credits have historically applied nationally, to estimate these credits' efficacy, previous work has relied on analysis of Section 1603 Grants \citep{johnston_nonrefundable_2019, aldy_investment_2023} and indirect evidence such as state incentives \citep{metcalf_investment_2010} or natural gas price instruments \citep{bistline_inframarginal_2025}. 
Energy Communities, a new siting provision of the Inflation Reduction Act, provide upwards of 10 percentage point increases in the baseline Investment and Production Tax Credits to disadvantaged communities that were economically reliant on fossil fuels.
Renewable energy projects in Energy Communities and fossil fuel reliant communities faced similar economic conditions and tax rates, but projects in Energy Communities received larger renewable energy incentives. 
This study uses fossil fuel reliant communities (with baseline credit rates) as a control for Energy Communities (with increased credit rates), to test for real responses to the Investment and Production Tax Credits and the downstream impacts on political support for renewables.	

I first document that Energy Communities had 33\% more renewable investment, primarily driven by wind installations.
Investment begins as soon as four months after the passage of the Inflation Reduction Act, suggesting that these policies were able to target shovel ready projects.
Increases in investment led to 31\% greater renewable production than control counties by 2024.
As expected, additional production lags investment, beginning one year after the Inflation Reduction Act passes. 

To understand the magnitude and speed of these responses, I utilize an understudied dataset on 5-year planned generator installations to identify inframarginal projects.
Accounting for preplanned developments reduces the investment responses by two-thirds, but still yields a statistically and economically significant 11\% increase in renewable energy investment within Energy Communities.

Next, I study the downstream impacts of investments on political support for renewable energy.
Support for state-level minimum renewable energy requirements rises by 5 percentage points in Energy Communities relative to control counties after the Inflation Reduction Act, with no comparable increases in support for carbon emission or clean air and water regulations---the response is specific to renewables, not a generalized environmental shift.
The aggregate effect masks sharp heterogeneity.
Self-identified ``not-strong'' Republicans drive the response, with a 10.5 percentage point increase off a 50-percentage-point baseline (a 21\% relative increase); parents increase support by 13\% across party.
Within Republicans, the response is concentrated in the moderates of the party who are most likely to be persuadable.
Republican increases in support are also strongest in individuals with working-class incomes.
Parents are the other group who, unconditional on party, strongly increase support for renewable energy by roughly 8 percentage points (a 13\% relative increase). 
These results represent a positive feedback loop between renewable investments and political support for renewable subsidies.

I subsequently examine potential channels for how the Investment and Production Tax Credits garnered political support for renewable energy. 
Motivated by the heterogeneous political responses, I study two spillover channels: 1) labor markets and 2) public good provision.
Beginning with the labor market, Energy Communities experienced significantly higher wages, with construction wages increasing 7\% relative to control counties.
The average construction salary in these areas falls within the income bin of Republicans that had the largest increase in support for renewable energy.
To understand the increase in parental support, I look at whether these investments helped the provision of an important public good for parents, schools. 
Renewable energy investments can significant increase local tax revenues and, in turn, public good outlays \cite{meza_throckmorton_2022}. 

This paper complements an empirical literature on tax credit design and impacts within the energy sector. 
Most of this work has focused on demand-side credits \citep{revelt_mixed_1998,sallee_surprising_2011,mian_effects_2012,boomhower_credible_2014, berkouwer_credit_2022, ito_selection_2023, ida_choosing_2026}. 
Comparatively, little work has studied the supply side responses to renewable energy tax credits. 
\cite{metcalf_investment_2010} is a longitudinal study of the user cost of capital which relates the PTC to wind investment. 
\cite{johnston_nonrefundable_2019} and \cite{aldy_investment_2023} study the relative performance of investment and production incentives in the wind sector using Section 1603 Grants. 
More recently, \cite{bistline_inframarginal_2025} compares production responses to changes in locational marginal prices within the renewable energy sector, but does not directly study renewable tax credits. 
\cite{ashenfarb_place-based_2025} looks at planned investment in Energy Communities using interconnection queuing data and projections of emissions, however, does not study realized investment following the passage of the Inflation Reduction Act nor production of renewable energy. 
My results provide a new direct estimate of realized investment and production responses for the Investment and Production Tax Credits respectively. 

This paper also contributes to the literature on business responses to place-based tax policies \citep{greenstone_does_2008, feyrer_did_2011, wilson_fiscal_2012, busso_assessing_2013, kline_local_2014, gaubert_place-based_2025,slattery_bidding_2025}. 
Unlike previous federal place-based policies such as Empowerment Zones and Opportunity Zones which are industry agnostic, Energy Communities provide a new opportunity to study a place-based industrial policy focusing on the renewable energy sector. 
Renewable place-based policies are especially policy relevant given the trillions spent globally to spur local economic development with targeted renewable investments both within the United States (Inflation Reduction Act) and in the European Union (European Green Deal).

There is a nascent literature focusing on the economic benefits of place-based renewable policies. 
\cite{fabra_renewable_2024} and \cite{scheifele_not_2025} study the local economic impacts of renewable energy project entry in Spain and Brazil. 
I contribute to this literature by using a natural experiment to test both how well place-based policies can encourage real local renewable energy investments and create local economic spillovers. 
My findings demonstrate robust realized production and investment responses to renewable energy incentives with significant increases in local wages. 

Finally, there is an unsettled literature on the political economy of renewable investments focusing on renewable energy development. 
\cite{stokes_electoral_2016} finds an electoral backlash to costly wind projects in Canada, while \cite{urpelainen_electoral_2022} finds wind investment increases Democratic vote shares.
I contribute to this literature by quantifying the upstream channel of how support for renewable energy policies change in response to renewable energy investments.
My results also contrast a popular hypothesis that special interest groups can short-circuit positive policy feedback loops \citep{stokes_short_2020}.
I find that within a year of receiving renewable energy investments, communities increase their support for these policies. 

The rest of this paper is organized as follows. 
Section 2 provides policy context of the Investment and Production Tax Credits and Energy Communities. 
Section 3 introduces the data. 
Section 4 describes the empirical methodology and results for renewable energy production and investment. 
Section 5 discusses local spillovers of renewable energy investments. 
Section 6 concludes.

\section{Policy Context}
\subsection{The Investment Tax Credit and the Production Tax Credit}
The Investment Tax Credit and Production Tax Credit are the main federal incentives for all renewable energy investments, established by the Energy Act of 1978 and the Energy Policy Act of 1992 respectively. 
In the modern iteration of these incentives, investors choose between receiving credits for the Investment or Production Tax Credit at each of their renewable energy projects. 
If an investor elects to receive the Investment Tax Credit for a project, they can claim a percentage of the project's investment cost as a tax credit when the project is put into service (begins selling electricity). 
If an investor elects to receive the Production Tax Credit, they can claim an annual tax credit per unit (kilowatt-hour, kWh) of electricity produced within a project's first ten years of service. 
Typically, wind energy has been the primary claimer of the Production Tax Credit, while solar energy has been the primary claimer of the Investment Tax Credit.

The Inflation Reduction Act (IRA) makes numerous adjustments to the Investment and Production Tax Credits. 
For tax credit rates, the IRA not only increased the maximum base rates for the Investment and Production Tax Credits, but also added domestic content and siting credit bonuses. 
The IRA increased the Investment Tax Credit's maximum base rate from 26\% to 30\%, and the Production Tax Credit's maximum base rate from 0.3 to 1.5 cents per kilowatt-hour. 
Each domestic content and siting bonus in the IRA adds 10 percentage points to the Investment Tax Credit base rates and 10\% of base credits to the Production Tax Credit. 
The siting bonuses in the IRA are the first geographical variation in Investment and Production Tax Credit incentives since 1978.

Under the IRA, the Investment Tax Credit or Production Tax Credit is available to wind, solar, geothermal electric, closed-loop biomass, open-loop biomass, landfill gas, municipal solid waste, and hydro. 
This set of energy sources will be referred to as renewable energy for the rest of this paper. 

\subsection{Energy Communities}
Energy Communities are a siting bonus for the Investment and Production Tax Credits established by the Inflation Reduction Act. 
This bonus is meant to encourage renewable energy investments in communities that are economically reliant on fossil fuels. 
There are three definitions of Energy Communities detailed in Table \ref{tab:ec_defs}.
The Statistical Area Category of Energy Communities is the focus of this paper.\footnote{The Brownfield Category of Energy Communities is not studied since it is defined at the property level, which requires individual micro data to study. The Coal Closure Category of Energy Communities is not studied since it receives pre-treatment renewable energy incentives from the Infrastructure Investment and Jobs Act of 2021.}
A Metropolitan Statistical Area (MSA) can be designated an Energy Community if its previous year unemployment rate was at least equal to the national average, and the MSA satisfies one of the following two criteria in any year since 2010: 1) 0.17\% of employment is directly related to coal, oil, or natural gas; or 2) at least 25\% of local tax revenues are related to coal, oil, or natural gas. 
The Treasury has not released public guidance on the local tax revenue eligibility criteria for Energy Communities, so, in practice, Statistical Area Category eligibility is determined by fossil fuel employment shares and local unemployment rates. 

Being designated an Energy Community grants locales an increase in renewable energy tax incentives. 
Under the Inflation Reduction Act, if a project begins construction after 2022, it is eligible to receive the Energy Community PTC tax credit bonus for up to ten years after the project is placed in service. 
If an investor chooses to claim the Investment Tax Credit instead, they would receive a one-time Energy Community bonus credit when the project is placed in service. 
For projects which begin construction in 2022, firms can still receive the Production Tax Credit bonus for Energy Communities if their project comes in service after 2022, but the annual PTC bonus is contingent on Energy Community status of the project's locale each year.

\section{Data}
\subsection{Energy Investment, Production, and Prices}
I collect data on energy investments and production from the Energy Information Administration \citep{eia_opendata_2025}. 
Investment data comes from Form EIA 860. 
EIA 860 is a census of energy producers in the United States and contains annual information on energy production capacity at the generator-level for every large plant. 
Importantly, this data includes each plant's latitude and longitude, allowing me to link power plants to Energy Communities. 
Publicly available data from the EIA does not contain investment costs for generators, so I use potential energy production measures as my main investment outcome, specifically nameplate capacity. 
Nameplate capacity is a manufacturer rating of a generator's theoretical maximum electricity production. 
I use county-level aggregates of nameplate capacity as a measure of renewable energy capital. 
Energy generation data is collected from EIA Form 923 data which includes annual energy generation by generator for every large plant in the United States. 

Data on county-level locational marginal prices of electricity is collected from the ReWEP tool \citep{millstein_renewables_2025}\footnote{Thank you to Dev Millstein for generously sharing this data with me.}.
The locational marginal price of electricity is the cost to purchase or sell the next megawatt-hour (MWh) of electricity at a specific node of the power grid. 

\subsection{County Level Data}

I collect county-level firm demographics and employment from the Quarterly Census of Employment and Wages. 
State-level corporate tax rates are collected from the Tax Foundation. 
For county-level Energy Community status and fossil fuel employment status, I transcribed Internal Revenue Service (IRS) notices \citep{irs_notice_2023_29, irs_notice_2023_47}.

\subsection{Cooperative Election Survey}

Individual-level political attitudes are collected from the Cooperative Election Survey (CES, formerly the CCES), an annual probability-based survey of approximately 60,000 U.S. adults each year that records political attitudes, vote intentions, and demographic characteristics \citep{schaffner_ces_common_cumulative_2024}.
For variables collected consistently across waves, I use the CES Common Content Cumulative file \citep{schaffner_ces_common_cumulative_2024}.
For policy preference questions, including support for state renewable energy requirements, support for EPA carbon regulation, and support for stricter clean air and water enforcement, I use the CES Cumulative Policy Preferences file \citep{schaffner_ces_policy_cumulative_2024} for waves 2006--2021 and individual annual policy modules for waves 2022--2024 \citep{schaffner_ces_2022, schaffner_ces_2023, schaffner_ces_2024}.

\section{Renewable Energy Production and Investment}
In this section, I present the empirical results for Energy Communities impact on renewable energy production and investment. 
First, I describe the supply side elasticities of interest. 
Second, I discuss the selection of treatment and control counties used for the analysis. 
Third, I test whether Energy Community bonuses caused greater renewable energy production and investment in targeted counties. 
Fourth, I present results on the relative efficiency of renewable energy projects in Energy Communities.

\subsection{Supply Side Elasticities}
The main empirical analysis explores the responsiveness of production and investment to Investment and Production Tax Credit bonuses. 
I am primarily interested in the net-of-incentives elasticity. 
The net-of-incentives elasticity compares economic responses to how changes in the Investment and Production Tax Credits impact the overall cost/revenue of a project. 
This statistic can inform any policy which impacts the net economic incentive for renewable energy investment and production. 

For the Investment Tax Credit, the base rate is 30\%, for projects which meet the prevailing wage requirement, and 40\% for projects in Energy Communities. 
To calculate the net-of-incentives elasticity, I must account for how the Investment Tax Credit changes both the tax shield of investment (depreciation deductions) and the direct cost of investment (the corporate income tax credit). 
For this calculation, I follow the user cost of capital model from \cite{hall_tax_1967} and include an investment tax credit which is not fully deducted from the depreciation basis of a project, accounting for the 50\% Investment Tax Credit reduction in the depreciation basis of projects. 
Using this new user cost of capital calculation, I estimate that the Energy Community Investment Tax Credit bonus corresponds to an 18\% decrease in the user cost of capital. 
See Appendix A for details. 
Thus, to measure the Investment Tax Credit net-of-incentives elasticity, I will use an 18\% reduction in the user cost of capital.

For the Production Tax Credit, the base rate is \textcent 2.75 per kWh of electricity.\footnote{I take the PTC rate for wind, solar, geothermal, and closed loop biomass projects which satisfy the prevailing wage requirements.} 
Projects in Energy Communities receive a 10\% boost in their credits to \textcent 3.025. 
To calculate the net-of-incentives elasticity, I measure the Energy Community PTC bonus' impact on marginal revenue of production to be 4.1\%. 
See Appendix A for details. 
Thus, to estimate the net-of-incentives elasticity, I will compare the percent change in production to a 4.1\% change in marginal revenue. 

\subsection{Fossil Fuel Employment and Energy Community Counties}
This study uses Energy Community (EC) status to estimate supply side responses to the Investment and Production Tax Credits. 
A subset of counties in an Energy Community are the treatment group that receives greater Investment and Production Tax Credit rates. 
A subset of counties which are not part of an Energy Community are the control group. 
Although the IRA makes many changes to the Investment and Production Tax Credits beyond Energy Community bonuses, non-siting changes impact both EC counties and control counties equally. 
Nevertheless, non-Energy Community counties and Energy Communities counties may not be equally attractive to renewable energy investors. 
To address this concern, I restrict the analysis to compare counties which were eligible to become Energy Communities. 

The Energy Community Statistical Area Category definition denotes a set of counties which are eligible for Energy Community designation due to similar levels of economic reliance on fossil fuels. 
For the rest of this paper, this set of potential Statistical Area Category Energy Communities counties will be called \textit{fossil fuel employment} counties. 
The control group is the subset of fossil fuel employment counties which are not Energy Communities. 
The treatment group is the set of fossil fuel counties which are Energy Communities exclusively due to the Statistical Area Energy Community definition (having a higher than national average unemployment rate). 
I make the exclusive Statistical Area Energy Community Category refinement to remove counties which contain Census Tracts that satisfy the Coal Closure Category definition of Energy Communities. 
Coal Closure Category communities were tagged to receive renewable energy incentives in the Infrastructure Investment and Jobs Act in 2021 (a year before the IRA), and, thus, I drop these counties from the analysis. 

\begin{figure}[h!]
    \centering
    \includegraphics[width=\linewidth]{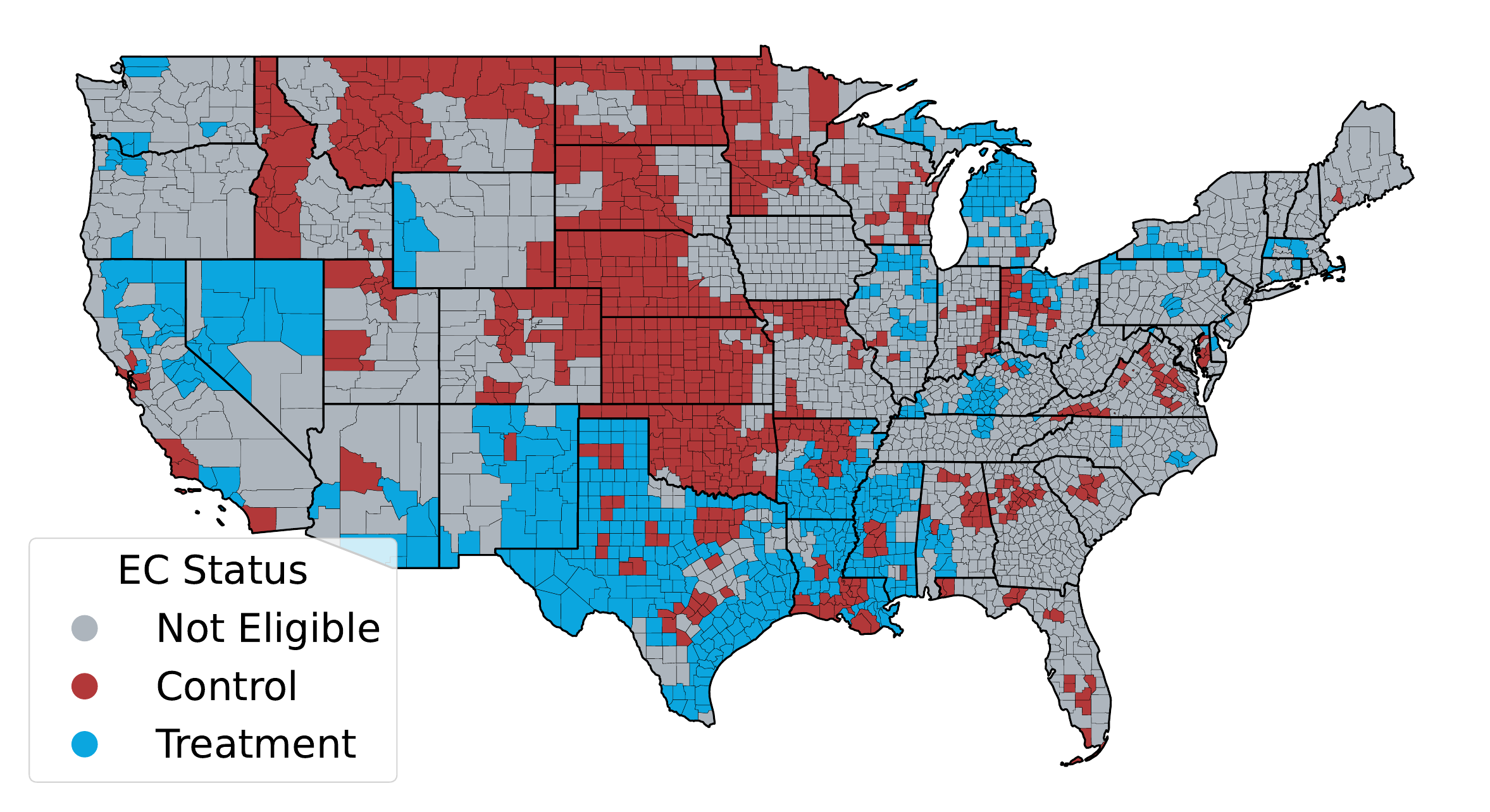}
    \caption{\textbf{Map of Energy Communities in 2023.} Orange and red counties have had at least 0.17\% direct employment in fossil fuel industries at least once since 2010. Red counties are designated Energy Communities in 2023 by satisfying both the direct employment criteria and having unemployment rates above the national average in 2022. \textit{Source: IRS.} }
    \label{fig:ec_map_2023}
\end{figure}

The final set of treatment counties is the set of Statistical Area Category Energy Communities which are not Coal Closure Category Energy Communities.
The final set of control counties is the set of fossil fuel employment counties which are not Energy Communities. 
For brevity, the rest of this paper will refer to Statistical Area Category Energy Communities as Energy Communities. 
Figure \ref{fig:ec_map_2023} highlights in red the treatment counties and in orange the control counties. 
Both treatment and control counties span across the United States with treatment counties being especially concentrated in the South and Midwest. 

\begin{figure}[h!]
    \centering
    \includegraphics[width=\linewidth]{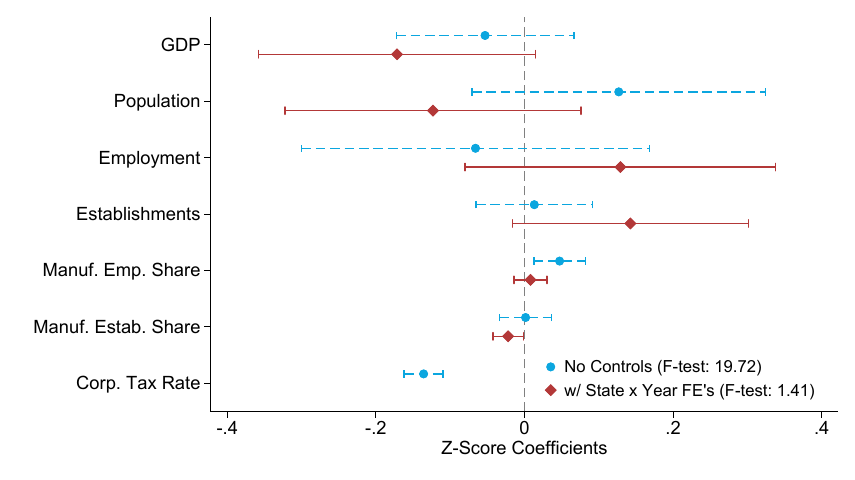}
    \caption{\textbf{Predictors of Treatment.} Regression coefficients of standardized Z-scores of county characteristics on a county's Energy Community status. \textit{Source: Tax Foundation, QCEW.} }
    \label{fig:trtmnt_pred}
\end{figure}

Table \ref{tab:ec_balance} provides summary statistics for the analysis sample of counties. 
Treatment and control counties have similar GDPs, populations, employed persons, establishments, and manufacturing establishments. 
However, treated counties appear to have lower corporate tax rates and higher manufacturing shares in employment. 
Figure \ref{fig:trtmnt_pred} shows the coefficients from regressing treatment on the standardized covariates in Table \ref{tab:ec_balance}. 
Lower state corporate tax rates and higher manufacturing shares predict treatment. 
I address this imbalance by controlling for county fixed effects and state-by-year fixed effects in the analysis.
County fixed effects control for permanent characteristics of counties such as renewable energy production potential.
State-by-year fixed effects control for any, observed or unobserved, shock to a state in a given year, such as changes to corporate tax rates or renewable energy portfolio standards \citep{feldman_renewable_2023}. 
In the analysis, the primary identifying variation comes from comparisons of EC counties to non-EC counties within a state and year. 

\subsection{Effect on Renewable Production and Investment}

\subsubsection{Renewable Production}
\begin{figure}[!ht]
    \centering
    \begin{subfigure}[b]{0.9\textwidth}
        \includegraphics[width=\linewidth]{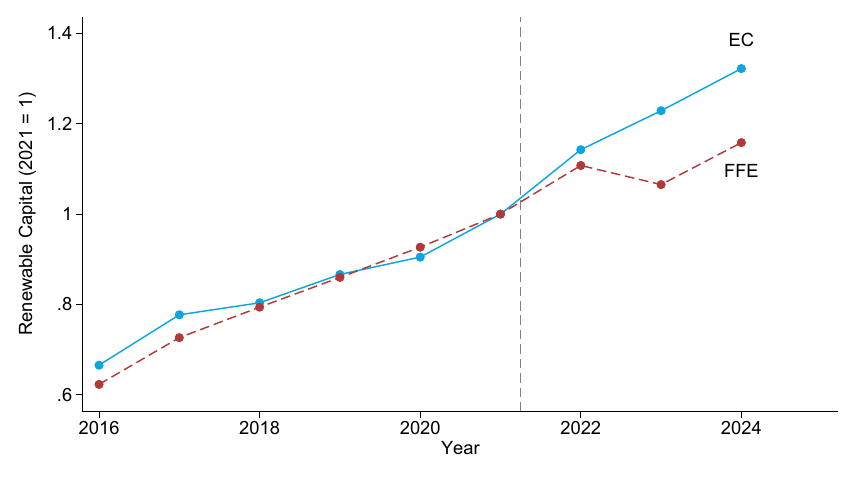}
        \caption{Generation}
        \label{fig:raw_mean_gen}
    \end{subfigure}
    \begin{subfigure}[b]{0.9\textwidth}
        \includegraphics[width=\linewidth]{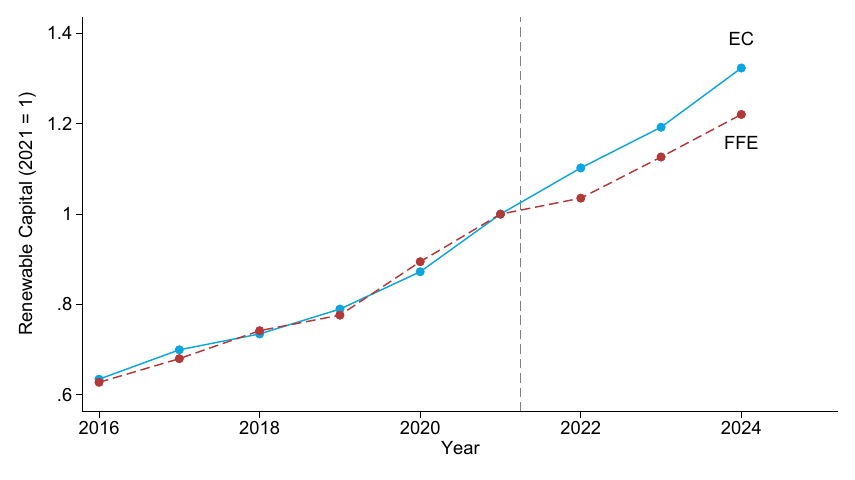}
        \caption{Investment}
        \label{fig:raw_mean_inv}
    \end{subfigure}
    \caption{\textbf{Impact of Energy Communities (Time Series).} Panel (a) plots the mean of renewable electricity generation capacity normalized such that 2021 is 1 by Energy Community status. 
    Panel (b) plots the mean of renewable nameplate capacity normalized such that 2021 is 1 by Energy Community status. 
    The dashed black line represents the passage of the Inflation Reduction Act in 2022.
    \textit{Source: EIA, IRS.} }
    \label{fig:raw_mean}
\end{figure}

Figure \ref{fig:raw_mean_gen} plots the time series of mean renewable energy generation, measured at the county-level, for Energy Communities (treatment) and fossil fuel employment counties (control), normalized to respective 2021 levels. 
The figure shows that the time series of normalized renewable energy generation in Energy Communities closely tracked normalized renewable generation in fossil fuel employment counties several years before 2022, suggesting the time series would have been similar in the absence of the IRA. 
Coincident with the passage of the IRA, normalized renewable generation in Energy Communities begins to outpace fossil fuel employment counties generation.
This figure provides suggestive evidence that Energy Community bonuses triggered real production responses within treated counties. 

To formalize this analysis, I conduct an event study estimation comparing renewable generation in treated counties and control counties around the passage of the Inflation Reduction Act. 
I use a Quasi-Maximum Likelihood Conditional Poisson model to avoid dropping zero outcome observations while using a log link between renewable outcomes and the independent variables. The estimating equation is: 

\begin{equation}
	\mathbb{E}[y_{c,t} \mid \cdot ] = \exp \left( \alpha + \sum_{k \in [-5,2] \setminus \{-1\}} \beta_k \cdot (EC_{c,t}^k) + \gamma_t + \lambda_c + \psi X_{c,t} \right) \epsilon_{c,t}
	\label{eq:did_poisson}
\end{equation}

\noindent where,  $y_{c,t}$ is an outcome in county $c$ in year $t$. $\alpha$ is the intercept. 
$\gamma_t$ are year fixed effects. 
$\lambda_c$ are county fixed effects. 
$EC_{c,t}^k$ is an indicator for if county $c$ is $k$ years from treatment in year $t$.
$X_{c,t}$ is a possibly empty vector of pre-period county characteristics interacted with year dummies. 
The coefficient $\beta_k$ captures the effect of Energy Community status in year $k$ relative to the pre-reform year, 2021. 
Standard errors are clustered at the level of treatment assignment, the MSA or non-MSA level. 

\begin{figure}[!ht]
    \centering
    \begin{subfigure}[b]{0.9\textwidth}
        \includegraphics[width=\linewidth]{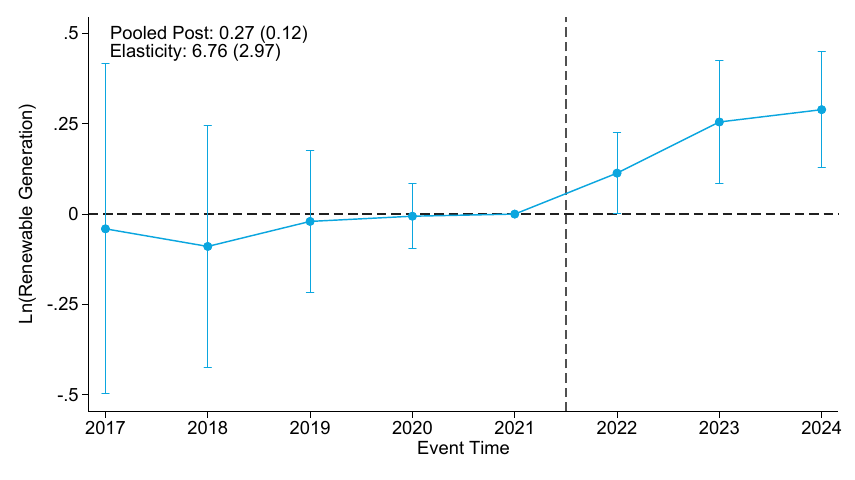}
        \caption{Generation}
        \label{fig:did_main_gen}
    \end{subfigure}
    \begin{subfigure}[b]{0.9\textwidth}
        \includegraphics[width=\linewidth]{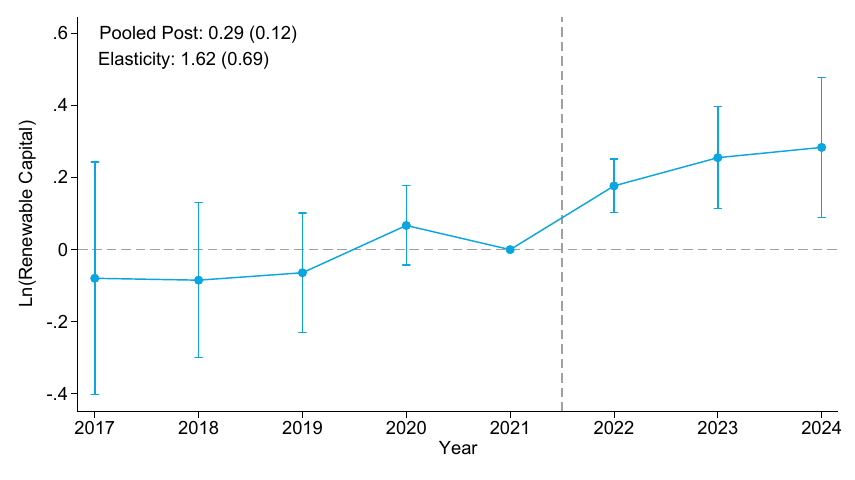}
        \caption{Investment}
        \label{fig:did_main_inv}
    \end{subfigure}
    \caption{\textbf{Impact of Energy Communities on Renewable Energy.} 
     Panel (a) plots the event study coefficients for Energy Communities' impact on renewable energy production.
     Panel (b) plots the event study coefficients for Energy Communities' impact on renewable energy investment. 
     Controls include state-by-year fixed effects and a lag of counties' locational marginal electricity price quartile (set to the 2021 value post IRA).
     The dashed vertical line represents the passage of the Inflation Reduction Act in 2022.
     \textit{Source: EIA, IRS.} }
    \label{fig:did_main}
\end{figure}

The key identifying assumption for this analysis is not that treatment and control counties were randomly assigned, but rather that renewable production in Energy Communities and fossil fuel employment communities would have trended similarly in the absence of the IRA, also known as the classic parallel-trends assumption. 
This assumption is supported by Figure \ref{fig:raw_mean_gen}, and can further be assessed by looking at the pre-trends from estimates of Equation \ref{eq:did_poisson}. 
Figure \ref{fig:did_main_gen} presents estimates for $\beta_t$ from Equation \ref{eq:did_poisson} where $X_{c,t}$ is a set of state-by-year fixed effects and a lag of counties' locational marginal electricity price quartile (set to the 2021 value post IRA). 
The event study figure shows that before 2023, there are no statistically significant differences in renewable generation between Energy Communities and fossil fuel employment communities, further supporting the identifying assumption. 

In Figure \ref{fig:did_main_gen}, Energy Communities begin producing more renewable electricity at a statistically significant level in 2023, one year after the IRA's passage. 
Table \ref{tab:main_did} Panel A reports the Difference-in-Difference estimates of the event study where the post period is aggregated into one event.  
Column 1 shows the baseline specification when $X_{c,t}$ is a set of state-by-year fixed effects. 
Column 2 presents the preferred specification, which adds a lag of the quartile of counties' locational marginal electricity price (assigned 2021 value for observations post IRA) to the baseline analysis. 
Column 2 of Table \ref{tab:main_did} reports that Energy Community counties produced, in expectation, 31\% ($=\exp(0.27)-1$) more renewable energy than fossil fuel employment counties in 2023. 
In levels this represents roughly 159,300 MWh of additional renewable energy which is both economically significant and statistically significant at the 5\% level. 
To contextualize this result, we can use the EIA's estimate of an average household's electricity consumption, 10,791 kWh. 
Thus, the renewable energy production response in a treatment county could power 14,762 ($=159,300 / 10,791 \times 1000$) homes for one year. 
On average there are about 41,030 occupied homes in each U.S. county, thus, the production response in treated counties is roughly 35\% ($= 14,762 / 41,030$) of annual residential energy consumption in each treated county. 
Generation responses are concentrated in 2023 and 2024. 
This is in line with the incentive structure of the Energy Community bonuses which only apply to electricity generated after 2022.

The EC bonus increased the marginal revenue of production by roughly 4.1 percent. 
This translates to a net-of-incentives elasticity of production of 6.11.

\subsubsection{Renewable Net Investment}
Figure \ref{fig:raw_mean_inv} replicates Figure \ref{fig:raw_mean_gen} for renewable investment. 
The time series present trends in county-level measures of renewable nameplate capacity, normalized to 2021 levels within Energy Communities and fossil fuel employment counties. 
The figure shows that immediately after the passage of the IRA, Energy Communities increase investment in renewable energy relative to fossil fuel employment counties.  
Table \ref{tab:main_did} Panel B replicates the analysis of Table \ref{tab:main_did} Panel A for renewable investment. 
Column 2 presents the preferred specification with state-by-year fixed effects and a lag of counties' locational marginal electricity price quartile.
Figure \ref{fig:did_main_inv} presents estimates for $\beta_t$ for the preferred specification of Equation \ref{eq:did_poisson}. 

The point estimate of the Difference-in-Difference coefficient, all post periods pooled into one event, for renewable nameplate capacity within Energy Communities compared to control counties is roughly 33\%, corresponding to an additional investment of 82 MW. 
To understand the economic significance of this estimate, we can use the EIA's 2022 estimate of the cost of installing a kilowatt of nameplate capacity, roughly \$1000. 
Thus, Energy Communities received, on average, an additional \$82 million ($\approx 82 \times 1000 \times 1000$) in renewable energy capacity. 
Assuming an 18\% reduction in the user cost of capital in Energy Communities, the corresponding net-of-incentives elasticity is 1.62.

\begin{figure}[!ht]
    \centering
    \includegraphics[width=0.9\linewidth]{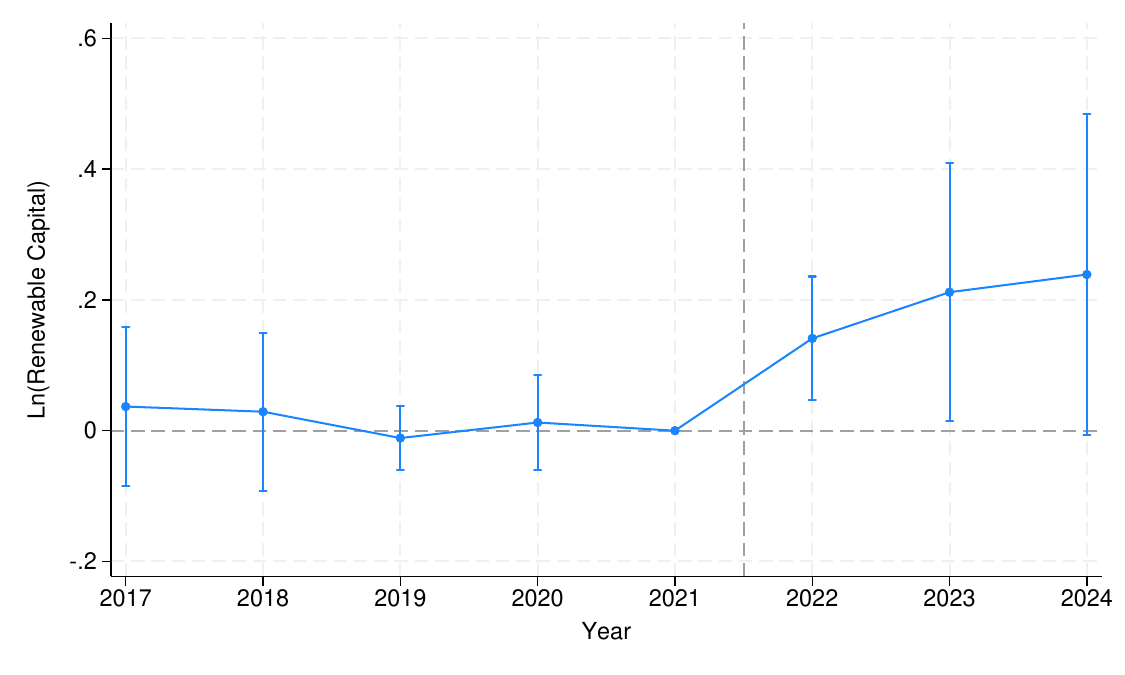}
    \caption{\textbf{Impact of Energy Communities on Renewable Energy (Unplanned).}
     Plots the event study coefficients for Energy Communities' impact on renewable energy investment less inframarginal projects.
     An inframarginal project is a project that was planned before the Inflation Reduction Act.
    \textit{Source: EIA, IRS.} }
    \label{fig:did_inframarginal}
\end{figure}

Adjusting for projects which were planned before the Inflation Reduction Act reduces this effect significantly.
Figure \ref{fig:did_inframarginal} presents estimates for $\beta_t$ from Equation \ref{eq:did_poisson} where the outcome is investment less projects planned before the Inflation Reduction Act.
While point estimates of the event study are similar, the initial magnitude of unplanned nameplate capacity in 2021 is much lower.
Thus, the unplanned investment response is only a 27 MW increase. 
Applying the same base capital stock as in the previous analysis yields a total investment increase of 11\% and an investment elasticity of 0.6.

\subsubsection{The Speed of Renewable Investment}
\begin{figure}[ht]
    \centering
    \includegraphics[width=\linewidth]{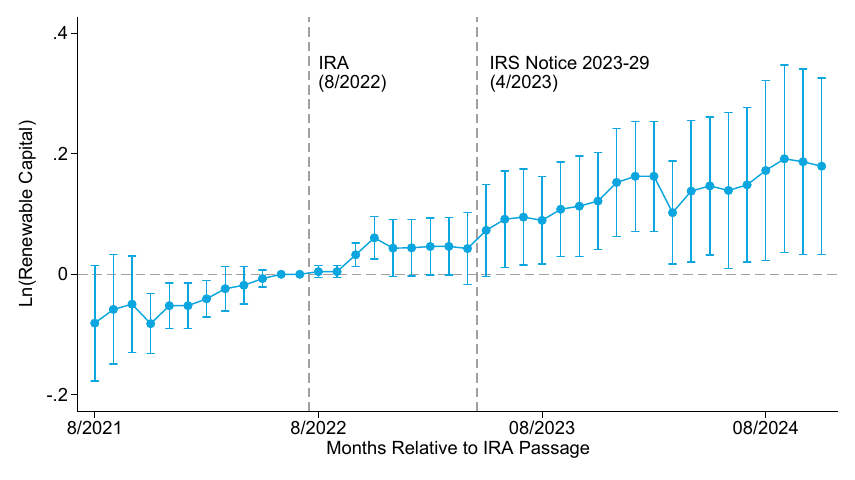}
    \caption{\textbf{Impact of Energy Communities on Renewable Investment (Monthly).} Plots the event study coefficients for Energy Communities' impact on renewable energy investment at the monthly level.
    \textit{Source: EIA, IRS.}}
    \label{fig:did_monthly_inv}
\end{figure}

Renewable energy investment increases in 2022, the same year as the IRA's passage.
One concern is that annual level measurements mask pre-trends in investment earlier in 2022, before the passage of the IRA. 
To address this, I replicate the investment analysis using a more granular time series of renewable energy investment, at the monthly level.
Figure \ref{fig:did_monthly_inv} presents estimates for $\beta_t$ from Equation \ref{eq:did_poisson} where $t$ is now a month-year combination.
The event study estimates show that the increase in renewable energy investment in Energy Communities in 2022 is concentrated a few months after the passage of the IRA. 
This suggests that the documented investment responses in 2022 are not driven by pre-trends, but rather by the increased credits offered by Energy Community status.

\subsection{Energy Community Relative Efficiency}
\begin{figure}[ht]
    \centering
    \includegraphics[width=\linewidth]{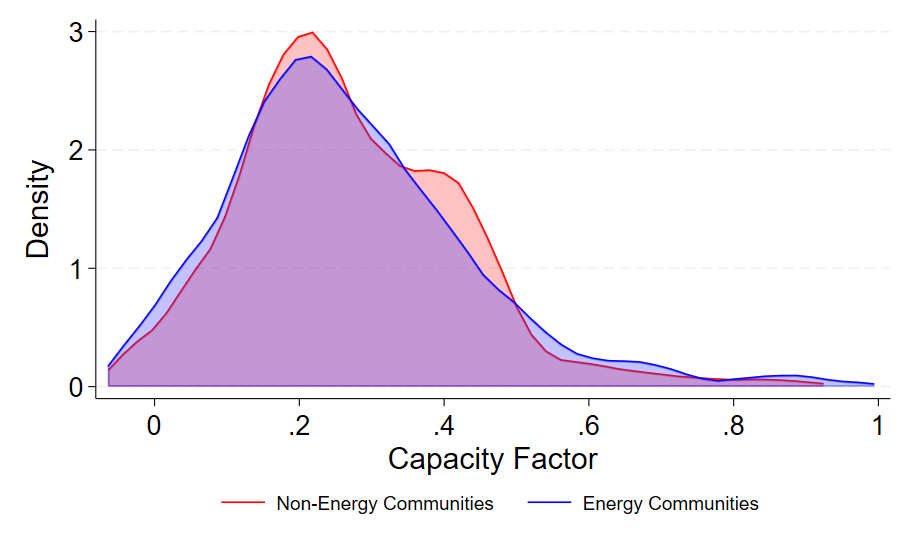}
    \caption{\textbf{Capacity Factor Comparison.} This figure plots the distribution of generators' capacity factors in Energy Communities and the rest of the mainland United States. 
    The capacity factor is the ratio of renewable energy generation to nameplate capacity.
    The mean capacity factor for Energy Communities and non-Energy Communities is 0.281. \textit{Source: IRS, EIA.}}
    \label{fig:capacity_factor}
\end{figure}

While the previous section shows that Energy Communities had greater renewable energy production and investment, investments in these counties are potentially less efficient than investments in other counties.
To test for this possibility, I compare the ratio of renewable energy production to nameplate capacity in Energy Communities and non-Energy Community counties. 
This ratio is a measure of the efficiency of renewable energy projects known as the capacity factor of a generator. 
Figure \ref{fig:capacity_factor} plots the distribution of capacity factors for renewable energy projects in Energy Communities and the rest of the mainland United States in 2021.
In both Energy Communities and control counties, the average capacity factor of renewable energy projects is roughly 0.28, suggesting that Energy Community projects are not less efficient than projects in control counties.

Another measure of efficiency is whether projects are being directed to where energy is most valuable. 
To test for this, I compare the locational marginal price (LMP) of electricity in Energy Communities and control counties in 2021. 
The LMP is the price of electricity at a specific location and time. 
In 2021, the average LMP in Energy Communities was \$33.81/MWh, while the average LMP in control counties was \$38.13/MWh, however this price difference is not statistically significant.

In terms of capacity factor and local electricity prices, Energy Community counties do not appear to be significantly less efficient locations for renewable energy projects than the rest of the mainland United States.

\section{Impact on Local Communities}
A goal of the Inflation Reduction Act's Energy Community bonuses was to create local economic benefits in communities reliant on fossil fuels.
The Biden administration hoped that this could create both an equitable green transition for fossil fuel workers and a political constituency for climate action.
These redistributive goals are not unique to the United States, as the European Union's Just Transition Fund has similar goals of creating local economic benefits in disadvantaged communities through renewable investments.

In this section, I test whether the tens of millions of dollars in renewable energy investments in Energy Communities created local economic benefits and whether these benefits translated into increased political support for renewable energy.
To test for economic benefits, I look at changes in local employment and wages in Energy Communities relative to control counties.
For political support, I look at changes in support for environmental policies using survey data from the Cooperative Election Survey.

\subsection{Political Impacts}
\begin{figure}[ht]
    \centering
    \includegraphics[width=\linewidth]{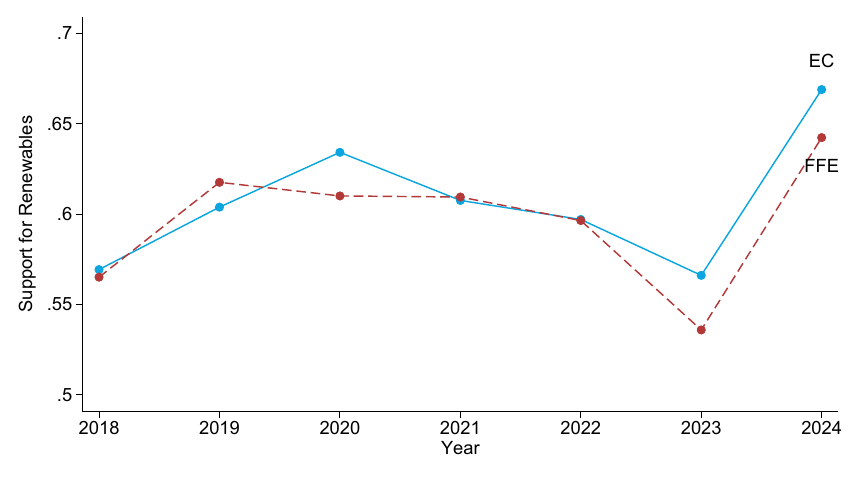}
    \caption{\textbf{Energy Communities and Renewable Energy Support (Time Series).} Plots of the mean of renewable energy support in Energy Communities (blue) and Fossil Fuel Employment counties (red).
    The point values for each term are from estimates of a regression of support for renewable energy on an Energy Community indicator and a control for self-identified political affiliation within each year. 
    \textit{Source: EIA, IRS.} }
    \label{fig:renewsup_trends}
\end{figure}

To test for Energy Communities' impact on political support for renewable energy, I use survey data from the Cooperative Election Survey (CES).
Figure \ref{fig:renewsup_trends} plots the time series of support for renewable energy across Energy Communities and control areas, controlling for self-disclosed political affiliation.
In 2022, residents in Energy Communities and control counties have, on average, nearly identical levels of support for renewable energy policies\footnote{The Cooperative Election Survey asks respondents if they support a minimum level of renewable energy share in their state. In 2021, the exact question was: ``Require that each state use a minimum amount of renewable fuels (wind, solar, and hydroelectric) in the generation of electricity even if electricity prices increase a little.''}.
In 2023 and 2024, support for renewable energy policies increases in Energy Communities relative to control counties.

Unlike investment, which increased by the end of 2022, CES estimates suggest that political support for renewable energy policies in Energy Communities did not increase until 2023.
Part of this is due to the survey timing of the CES which collected political support for environmental policies from respondents in September and October of 2022. 
Comparing this to the monthly investment responses in Figure \ref{fig:did_monthly_inv}, CES responses were collected before most of the renewable energy responses to the IRA.
As such, for the rest of this section, I will treat 2023 and 2024 as the post period for analyzing how the IRA's Investment and Production Tax Credit incentives impact political support for renewable energy policies.

I use a Difference-in-Difference approach to compare support for different environmental policies in Energy Communities relative to control counties around the passage of the IRA. 
The estimating equation is:
\begin{equation}
    y_{ipt} =\alpha + \beta ( EC_{m(i)} \times \text{Post}_t) + \lambda_{m(i)} + \psi X_{s(i)t} + \epsilon_{i,t}
    \label{eq:dd_climate_support}
\end{equation}
\noindent where,  $y_{ipt}$ is a dummy for individual $i$'s support of environmental policy $p$ in year $t$. 
$\alpha$ is the intercept.
$\lambda_{m(i)}$ are MSA fixed effects.
$EC_{m(i)}$ is the treatment status of MSA $m(i)$, and $\text{Post}_t$ is a dummy for whether year $t$ is 2023 or later. 
$X_{c,t}$ is a set of state-by-year fixed effects and county Republican vote shares in the last presidential election. 
The coefficient $\beta$ captures the average effect of Energy Community status on support for environmental policy $p$ after the IRA relative to before.
The key identifying assumption for this analysis is that, in the absence of the IRA, support for environmental policies would have trended similarly in Energy Communities and control counties. 
The similarities in support before the IRA from Figure \ref{fig:renewsup_trends} support this assumption.  

\begin{figure}[ht]
    \centering
    \includegraphics[width=\linewidth]{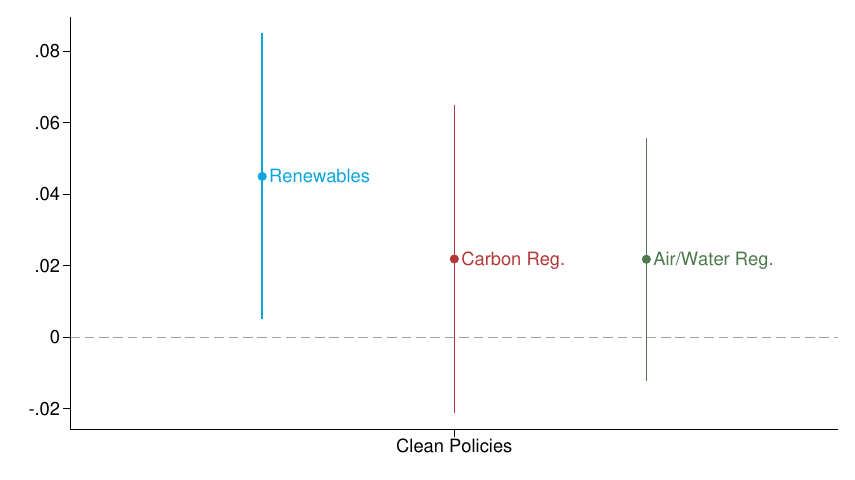}
    \caption{\textbf{Impact of Energy Communities on Support For Environmental Policies.} Plots the Difference-in-Difference coefficient for Energy Communities' impact on support for three sets of environmental policies.
        ``Renewables'' refers to support for policies requiring a minimum amount of renewable energy.
        ``Carbon Ref.'' refers to support for the EPA regulating carbon emissions. 
        ``Air/Water Reg.'' refers to support for more strictly enforcing clean air and water regulation.
        Controls include a set of MSA and state-by-year fixed effects.
    \textit{Source: CES, IRS.} }
    \label{fig:enviro_policies_did}
\end{figure}

Figure \ref{fig:enviro_policies_did} presents estimates for $\beta$ from Equation \ref{eq:dd_climate_support} for three sets of policies: minimum renewable energy requirements, new carbon emission regulations, and stricter clean air and water regulations.
Energy Communities had a statistically significant 5 percentage point increase in support for minimum renewable energy requirements after the IRA relative to control counties, however, there was no significant change in support for new carbon emission regulations or stricter clean air and water regulations.
This suggests that the increase in renewable energy support among Energy Community residents is specific to renewable energy, and does not reflect a general increase in concerns about environmental issues.

While overall support for renewable energy policies increased in Energy Communities, an explicit goal of the IRA's Energy Community bonuses was to create a political constituency for climate action among Republicans within these historically fossil fuel-dependent communities.
To investigate this, I estimate a triple-difference regression along individuals' self-disclosed political affiliation.
The estimating equation is:
\begin{equation}
    \begin{split}
        y_{ipt} = & \alpha + \beta (EC_{m(i)} \times \text{Post}_t) + \beta^{\text{Trait}} (\text{Trait}_i \times EC_{m(i)} \times \text{Post}_t) \\
            & + \text{Trait}_i + (\text{Trait}_i \times \text{Post}_t) + (\text{Trait}_i \times \text{EC}_{m(i)}) + \lambda_{m(i)} + \psi X_{s(i)t} + \epsilon_{i,t}
    \end{split}
    \label{eq:ddd_renewsup}
\end{equation}      
\noindent where $\text{Trait}_i$ is a dummy for a self-disclosed categorical trait of individual $i$. All other terms are defined as in Equation \ref{eq:dd_climate_support}.
The key identifying assumption for this analysis is that, in the absence of the IRA, the gap in support for renewable energy policies across $\text{Trait}$ would have trended similarly in Energy Communities and control counties. 

\begin{figure}[ht]
    \centering
    \includegraphics[width=\linewidth]{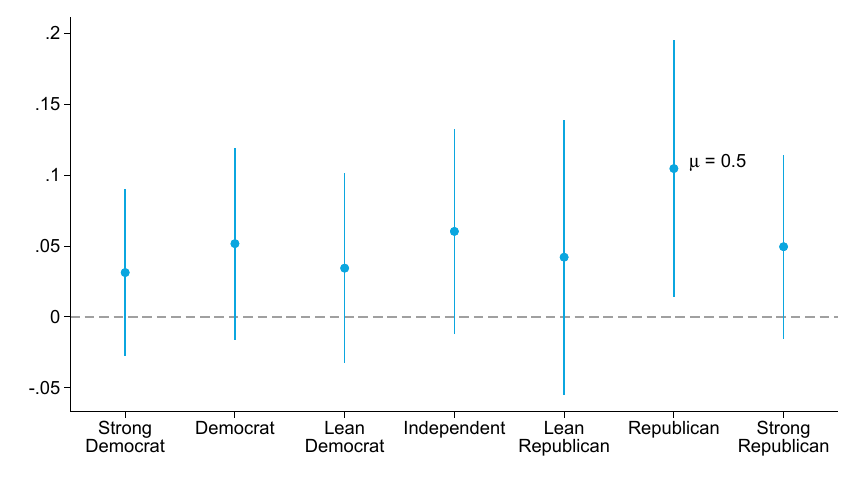}
    \caption{\textbf{Support For Environmental Policies Across the Political Spectrum.} Plots the triple-difference coefficients for Energy Communities' impact on support for renewable energy across political affiliation.
        The x-axis denotes self-disclosed political affiliation, $\text{Trait}_i$, of respondents in the Cooperative Election Survey.
        Each point estimate is the Energy Community $\times$ Post coefficient for the indicated affiliation group from Equation \ref{eq:ddd_renewsup}.
        Controls include a set of MSA and state-by-year fixed effects.
    \textit{Source: CES, IRS.} }
    \label{fig:renewsup_repub_ddd}
\end{figure}

Figure \ref{fig:renewsup_repub_ddd} presents estimates from Equation \ref{eq:ddd_renewsup} where $\text{Trait}_i$ is an individual's self-disclosed political affiliation in the CES survey.
This figure shows that the increase in support for renewable energy policies in Energy Communities was primarily driven by Republicans, with no other group demonstrating a statistically significant increase in support for renewable energy.
Further, this increase is large with point estimates suggesting a 10.5 percentage point increase in renewable energy support on a basis of 50 percentage points among Republicans in Energy Communities.
This represents a 21\% increase in support for renewable energy policies among Republicans in Energy Communities relative to Republicans in control counties. 
For Republicans, I can rule out increases in support as small as 5 percentage points with 95\% confidence.
This suggests that increased PTC and ITC incentives in Energy Communities successfully generated political support for renewable energy among Republicans in historically fossil fuel-reliant communities.

\subsubsection{Political Mechanisms}
To understand potential mechanisms for the increase in renewable energy support among Republicans in Energy Communities, I estimate two more triple-difference regressions across family income and parental status. 
Studying responses across family income provides suggestive evidence on whether the increase in renewable energy support among Republicans comes from individuals in the income distribution most likely to benefit from the documented increases in construction wages.
Studying responses across parental status provides suggestive evidence on whether the increase in renewable energy support among Republicans comes from individuals with children who may benefit from increased property tax revenues that could be targeted towards local public goods such as schools.
The estimating equation still follows Equation \ref{eq:ddd_renewsup}, but now $\text{Trait}_i$ is either a binned value of family income or a dummy for parental status.

\begin{figure}[ht]
    \centering
    \includegraphics[width=\linewidth]{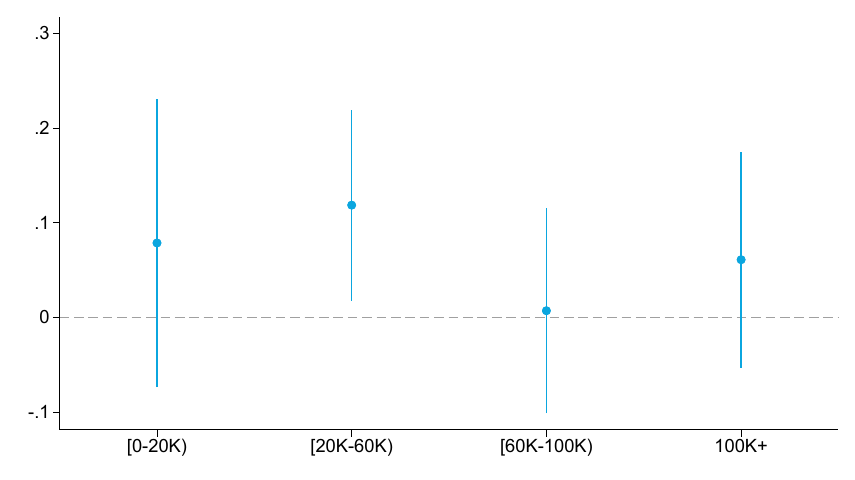}
    \caption{\textbf{Support For Environmental Policies Across the Income Distribution.} Plots the triple-difference coefficients for Energy Communities' impact on support for renewable energy across the political spectrum.
        The x-axis denotes the bins of family income, $\text{Trait}_i$, of respondents in the Cooperative Election Survey.
        The $[0-\$20K)$ bin estimate is from  $\beta$ from Equation \ref{eq:ddd_renewsup}.
        $\mu$ is the mean level of renewable energy policy support among Energy Communities' Republicans in 2022.
        Controls include a set of MSA and state-by-year fixed effects.
        \textit{Source: CES, IRS.}}
    \label{fig:repub_faminc_renewsup_ddd}
\end{figure}

\textit{Labor Market.} Figure \ref{fig:repub_faminc_renewsup_ddd} presents the triple-difference estimates from Equation \ref{eq:ddd_renewsup} where $\text{Trait}_i$ is a bin of self-reported family income.
I restrict this analysis to Republicans, to disentangle what drives their increase in renewable energy support.
The figure shows that the increase in renewable energy support among working-class Republicans in Energy Communities is primarily driven by individuals with family incomes between \$20,000 and \$60,000. 

I test for this channel in the Energy Communities' local labor markets. 
Since Energy Communities were targeted towards high unemployment rate areas, a single event study framework such as Equation \ref{eq:did_poisson} will not be sufficient to test for changes in local employment and wages.
Instead, I use a triple difference approach to compare changes in employment and wages in the construction industry (the industry most likely to be impacted by renewable energy investments) relative to control industries across Energy Community and control counties.
I estimate this triple difference using a Quasi-Maximum Likelihood Conditional Poisson model, and the estimating equation is:
\begin{equation}
\begin{split}
    \mathbb{E}[y_{i,c,t} \mid \cdot ] & = \exp \left(\alpha + \sum_{t \not = 2021 } \beta_t^1 ( \text{Treat}_i \times \text{Year}_t) 
           + \sum_{t \not = 2021 } \beta_t^2 ( EC_c \times \text{Year}_t) \right. \\
           & + \sum_{t \not = 2021 } \beta_t^3 ( EC_c \times \text{Year}_t \times \text{Treat}_i) \left. + \gamma_t + \lambda_c + \eta_{i} + \psi X_{c,t} \vphantom{\sum_{t \not = 2021 } \beta_t^1 ( \text{Treat}_i \times \text{Year}_t)} \right) \epsilon_{i,c,t}
\end{split}
\label{eq:ddd_emp_wages}
\end{equation}

\noindent where,  $y_{i,c,t}$ is an outcome for industry $i$ in county $c$ in year $t$. 
$\alpha$ is the intercept. 
$\gamma_t$ are year fixed effects. 
$\lambda_c$ are county fixed effects. 
$\eta_i$ are industry fixed effects. 
$\text{Treat}_i$ is a dummy variable indicating if industry $i$ corresponds to construction. 
$EC_c$ is the treatment status of county $c$, and $\text{Year}_t$ are dummies for each year. 
$X_{c,t}$ is a set of state-by-year fixed effects.
Standard errors are clustered at the level of treatment assignment, the MSA or non-MSA level.
$\beta^3_t$ captures the difference in the treated and non-treated industry gap between Energy Communities and control counties.

\begin{figure}[ht]
    \centering
    \begin{subfigure}[b]{0.45\textwidth}
        \includegraphics[width=\linewidth]{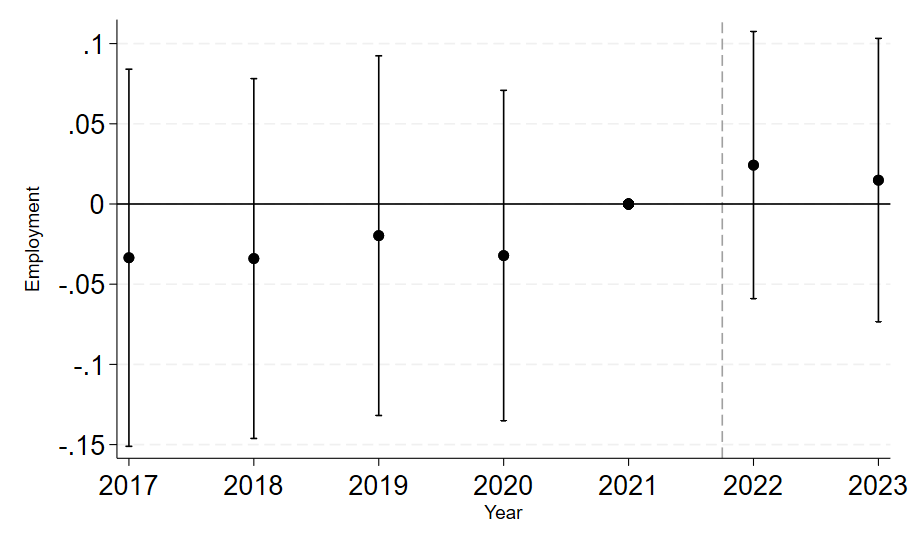}
        \caption{Employment}
        \label{fig:ddd_emp}
    \end{subfigure}
    \hfill
    \begin{subfigure}[b]{0.45\textwidth}
        \includegraphics[width=\linewidth]{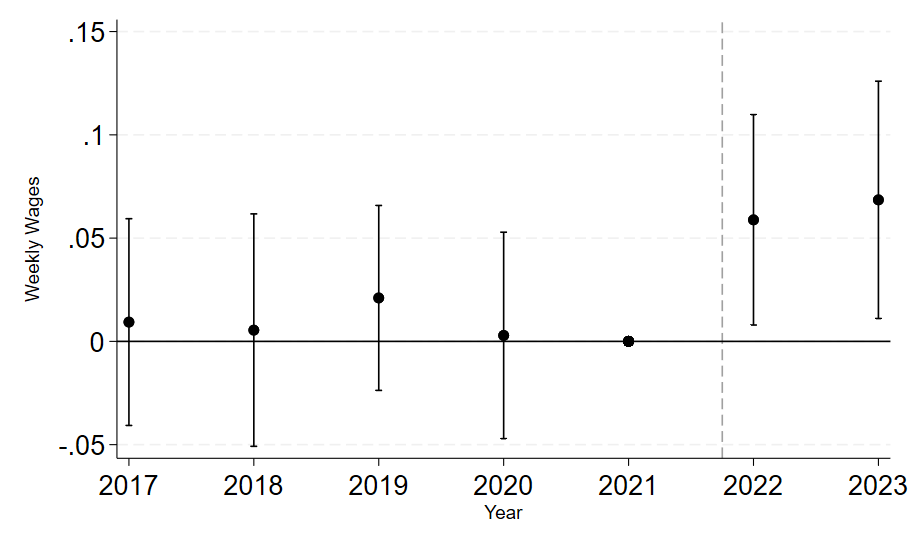}
        \caption{Weekly Wages}
        \label{fig:ddd_wages}
    \end{subfigure}
    \caption{\textbf{Local Economic Spillovers of Energy Communities.} 
     Panel (a) plots the event study coefficients for Energy Communities' impact on construction employment relative to control counties.
     Panel (b) plots the event study coefficients for Energy Communities' impact on construction weekly wages relative to control counties. 
     Controls include state-by-year fixed effects and a lag of counties' locational marginal electricity price quartile (set to the 2021 value post IRA).
     The vertical axis is the natural log of employment and weekly wages respectively.
     The dashed vertical line represents the passage of the Inflation Reduction Act in 2022.
     \textit{Source: IRS, QCEW.} }
    \label{fig:ddd_emp_wages}
\end{figure}

The key identifying assumption for this analysis is that, in the absence of the IRA, the difference in employment and wages between construction and control industries would have trended similarly in Energy Communities and control counties.
Figure \ref{fig:ddd_emp} and Figure \ref{fig:ddd_wages} present estimates for $\beta^3_t$ from Equation \ref{eq:ddd_emp_wages} for employment and wages respectively.
The figures show that construction employment and wages in Energy Communities trended similarly to control counties before 2022.
After the passage of the IRA, construction wages in Energy Communities increased by roughly 7\% relative to control counties, however, there is no statistically significant change in construction employment.
This provides suggestive evidence that the increase in renewable energy investments in Energy Communities created local economic benefits through increased wages for construction workers, which could be a channel for the increase in renewable energy support among working-class Republicans in Energy Communities.

\begin{figure}[ht]
    \centering
    \includegraphics[width=\linewidth]{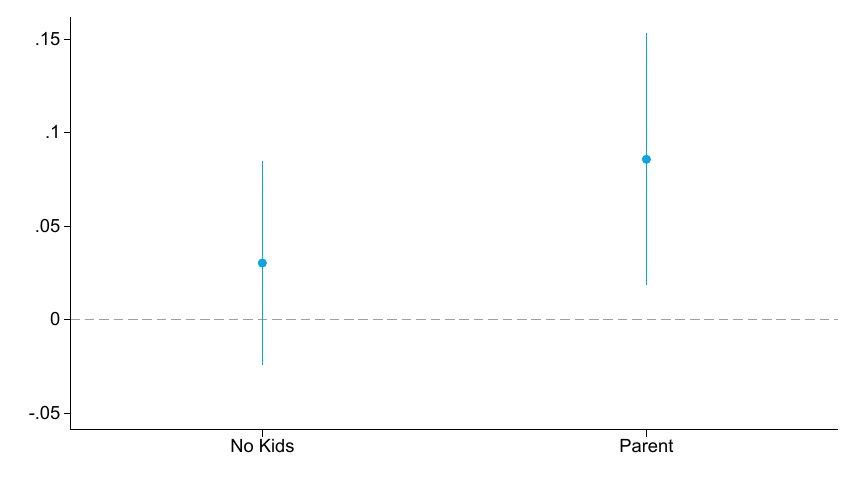}
    \caption{\textbf{Support For Environmental Policies Across Parental Status.} Plots the triple-difference coefficients for Energy Communities' impact on support for renewable energy across the political spectrum.
        The x-axis denotes parental status, $\text{Trait}_i$, of respondents in the Cooperative Election Survey.
        The ``No Kids'' estimate is from  $\beta$ from Equation \ref{eq:ddd_renewsup}.
        The ``Parent'' estimate is $\beta + \beta^{\text{Trait}}$ from Equation \ref{eq:ddd_renewsup}.
        Controls include a set of MSA and state-by-year fixed effects.
        \textit{Source: CES, IRS.}}
    \label{fig:parental_renewsup_ddd}
\end{figure}

\textit{Public Goods Provision.} Figure \ref{fig:parental_renewsup_ddd} presents estimates from Equation \ref{eq:ddd_renewsup} where $\text{Trait}_i$ is an individual's parental status.
Across all parents, regardless of political affiliation, there is a statistically significant increase in support for renewable energy policies in Energy Communities relative to control counties.
One channel for this differential effect among parents could be the provision of public goods, schools in particular. 
Anecdotally, testimonials from Energy Communities have cited increases in property tax revenues from renewable energy projects as key drivers for construction of new school facilities \citep{meza_throckmorton_2022}.

\section{Conclusion}
In the United States, energy production is the second most polluting industry in terms of greenhouse gas emissions.
Due to the political disaffection towards a federal carbon tax, the United States has relied on tax credits to encourage clean energy investments and production.
For nearly 50 years, the Production Tax Credit and Investment Tax Credit have been the two largest green subsidies in the United States. 
Despite their importance, there is still little evidence on the magnitude of supply side responses to these incentives.

This paper uses the Inflation Reduction Act's Energy Community bonuses to estimate the supply side responses to the PTC and ITC. 
I find that Energy Communities had 33\% greater renewable energy investment and 31\% more renewable production than control counties after the passage of the IRA.
The PTC and ITC have net-of-incentives elasticities of 1.62 and 6.11 respectively.
However, accounting for inframarginal projects significantly reduces investment elasticities to 0.6.
These findings are policy relevant for the design of renewable energy incentives in the United States and globally, as governments are spending trillions of dollars to encourage renewable energy investments.

Further, these incentives create a positive policy feedback loop for renewable energy policies.
Areas that received greater renewable energy investments and production due to the IRA's Energy Community bonuses also had greater increases in support for renewable energy policies, especially among Republicans.
I demonstrate two potential channels for this increase in support among Republicans: labor market spillovers and increased public goods provision.
While construction jobs are short-lived, newly constructed renewable energy generators will be in communities for decades. 
This paper's analysis of Energy Community residents' political attitudes suggests mechanisms such as public goods provision can lead to long-term benefits for local residents.

\pagebreak

\printbibliography

\pagebreak

\section{Figures and Tables}

\begin{table}[h!]
\centering
\begin{tabular}{|p{0.25\textwidth}|p{0.6\textwidth}|}
\hline
Energy Community Type & Qualifications \\
\hline
Brownfield Category & Property that is complicated to develop due to a hazardous substance, pollution, or contaminant.  \\
\hline
Statistical Area Category & An MSA, or non-MSA, that has: \newline 1. at any time since December 31, 2009, had 0.17 percent or greater direct employment in fossil fuel industries or at least 25 percent or greater local tax revenues related to fossil fuel industries, and \newline 2. an unemployment rate at least equal to the national average \\
\hline
Coal Closure Category & A census tract, and directly adjoining tract, in which either \newline 1. a coal mine has closed since December 31, 1999, or \newline 2. in which a coal-fired electric generating unit has retired after December 31, 2009 \\
\hline
\end{tabular}
\caption{\textbf{Energy Community Definitions.} Definitions of the three categories of Energy Communities defined by the Inflation Reduction Act. 
Only Statistical Area Category Energy Communities are included in the main analysis. 
\textit{Source: IRS.}}
\label{tab:ec_defs}
\end{table}

\pagebreak

\begin{table}[h!]
    \centering
    \begin{tabular}{l|cc}
	        \hline
        & Control & Treatment  \\
        \hline
        GDP (Billion \$) & 5.06 & 6.88\\
         & (22.34) & (45.06) \\
         Population (1,000s) &  67.35 & 95.64\\
         &    (219.78) &  (496.85) \\
        Total Emp. (1,000s) &       24.88 &       34.26\\
        &     (98.52) &    (202.68) \\
        Total Estab. (1,000s)&        1.90 &        2.89\\
        &      (7.10) &     (22.20) \\
        Manuf. Emp. Share (\%)&       12.65 &       13.93\\
        &     (12.19) &     (12.85) \\
        Manuf. Estab. Share (\%)&        4.28 &        4.46\\
        &      (2.28) &      (2.42) \\
        Corp. Tax Rate (\%) &        5.73 &        4.14\\
        &      (2.58) &      (3.40) \\
        \hline
        Observations        &         746        &         585            \\
        \hline
    \end{tabular}
    \caption{\textbf{Fossil Fuel Employment and Energy Community Counties Balance.}
    Summary statistics for the analysis sample of counties.
    Control counties are fossil fuel employment counties which are not Energy Communities.
    Treatment counties are Statistical Area Category Energy Communities which are not Coal Closure Category Energy Communities.
    \textit{Source: Tax Foundation, QCEW.}}
	\label{tab:ec_balance}
\end{table}

\pagebreak
\begin{table}[h!]
\centering
\begin{tabular}{lcc}
    \toprule
    & (1) & (2) \\
    \midrule
    \multicolumn{3}{l}{\textbf{Panel A. Investment}} \\
    \addlinespace
    $EC_c \times Post$		 & 0.25 & 0.29  \\
                         & (0.13) & (0.12)  \\
    \addlinespace
    \midrule
    N & 8,520 & 5,701 \\
    \addlinespace
    EC Mean 2021 &  192 MW & 248 MW \\
    \addlinespace
    Implied $\epsilon$ wrt (1 - $\tau$)    & 1.39 &  1.62 \\
    & (.71) & (0.69) \\
    \addlinespace

    \bottomrule
    \addlinespace
    \multicolumn{3}{l}{\textbf{Panel B. Production}} \\
    \addlinespace
    $EC_c \times Post$			 & 0.26  & 0.27  \\
                       & (0.12) & (0.12) \\
    \addlinespace
    \midrule
    N & 8,471 & 5,701 \\
    \addlinespace
    EC Mean 2021 & 466,712 MWh &  589,990 MWh \\
    \addlinespace
    Implied $\epsilon$ wrt (1 - $\tau$) &  6.26  & 6.11 \\
    & (2.90) & (2.71) \\
    \midrule
    
    $State_c \times Year_t$       & Yes & Yes \\
    $LMP_{c,t-1}$       & No & Yes \\      
    \bottomrule
\end{tabular}
\caption{\textbf{Effect of Energy Communities on Renewable Energy Production and Investment.}
Estimates of the effect of Energy Communities on renewable energy production and investment from the event study using estimating Equation \ref{eq:did_poisson}.
}
\label{tab:main_did}
\end{table}

\pagebreak

\pagebreak

\pagebreak

\pagebreak

\pagebreak

\pagebreak

\pagebreak

\pagebreak

\pagebreak

\pagebreak

\pagebreak

\pagebreak

\pagebreak

\pagebreak

\appendix

\section{Net of Incentives Elasticity Calculations}\label{app:elasticity}
This section describes the derivations of the net of incentives elasticity equations used in the analysis. While the policy elasticity is relevant for understanding how changes in the ITC and PTC impact economic behavior, the elasticity with respect to net incentives provides useful information on economic responses to any policy which targets the net economic incentive for renewable energy production or investment. 

\subsection{Production Tax Credit Net of Incentives Elasticity}
To get an equation for the Production Tax Credit elasticity, we begin with a stylized model of the renewable energy producer which maximizes the following profit equation: 
\begin{equation*}
	\pi = Q \cdot (p + c) \cdot (1 - t) - C(Q)
\end{equation*}
Where $Q$ is the quantity of electricity produced. $p$ is the price of electricity. $c$ is the production tax credit per unit of electricity. $t$ is the corporate tax rate. $C(Q)$ is a variable cost function for the production of $Q$ units of electricity. This profit equation treats the production tax credit as earned income. This follows the Inflation Reduction Act's changes to the PTC which made it transferable, so renewable energy producers could sell PTC credits to third parties to earn taxable income. 

Producers maximize profits by setting marginal profits to zero, or equalizing the marginal cost and marginal revenue of production: 
\begin{align*}
  	 \frac{\partial \pi}{\partial Q} = 0 &\implies \underbrace{(p + c) \cdot (1-t)}_{MR} = C'(Q)
\end{align*}
So the marginal net-of-tax revenue, $MR$, from an additional unit of production after taxes is given by $(p + c) \cdot (1-t)$. We are interested in the production elasticity to the marginal net-of-tax revenue, defined as 
\[
	\epsilon_{P} = \frac{dQ}{d(MR)} \frac{MR}{Q} \approx \frac{\Delta \log(Q)}{\Delta \log(MR)}
\]
\noindent where the second approximation is the estimable parameter of interest. The variation in marginal net-of-tax return within the analysis comes from Energy Community bonuses which increase the PTC credits from $c_0$, the base production tax credit, to $c_{EC}$. Substituting for the net-of-tax rate and including the source of variation:
\begin{align*}
\Delta \log(MR) & = \Delta \log \left((p + c) \cdot (1 - t) \right) \\
  &= \log \left((p + c_{EC}) \cdot (1 - t) \right) - \log \left((p + c_{0}) \cdot (1 - t) \right) \\
  &= \log (p + c_{EC}) - \log (p + c_0) \\
  &= \log \left( \frac{p + c_{EC}}{p+c_0}\right)
\end{align*}
The $1-t$ terms drop out since treatment and control communities only differ in their tax rates due to Energy Community bonuses. Thus our estimating equation for the production elasticity is 
\[
	\epsilon_{P} = \frac{\Delta \log(Q)}{\log \left( \frac{p + c_{EC}}{p+c_0}\right)}
\]
This expression is similar to a production elasticity with respect to a per unit production subsidy given the transferability of tax credits. 

To estimate $\epsilon_P$, $\Delta \log(Q)$ will be taken from analysis of Equation \ref{eq:did_poisson}. 
I take estimates of the price of wind generated electricity, $p$, from the average Power Purchase Agreements of \textcent 3.75 per kWh \citep{millstein_renewables_2025} in 2023.
For the baseline 2023 PTC rate, $c_0$, I use \textcent 2.75 per kWh of electricity.
Finally, following the 10\% Energy Community bonus to base credits, I use \textcent 3.025 for $c_{EC}$. 
Then, the Energy Community bonus leads to a $4.1\%$ change in after-tax revenue.

Two discrepancies between theory and estimates should be noted. First, \textcent 3.75 per kWh of electricity is the average price for 2023, not the marginal price. Second, the baseline credit, $c_0$ I am using is the maximum baseline rate received by projects which meet prevailing wage requirements. Further, the baseline credit I am using applies to wind, solar, geothermal, and closed-loop biomass projects.\footnote{Other renewable energy projects received \textcent 1.50 per kWh.} Readers can apply their own assumed tax changes to the raw estimates of production responses.

\subsection{Investment Tax Credit Net of Incentives Elasticity}
\subsubsection{The User Cost of Capital}
To get an equation for the Investment Tax Credit elasticity with respect to the user cost of capital, I follow the standard user cost of capital from \cite{hall_tax_1967} accounting for an investment tax credit which reduces the depreciation basis of investment by $\theta (< 1)$ of the credit, as opposed to the full amount. 
This adjustment leads to the following equation for the price of capital:
\[
	q(t) = \int_{t}^{\infty} e^{-r(s-t)} ((1-u)c(s)e^{-\delta(s-t)}+u(1-\theta k)q(t)D(s))ds + kq(t)
\]
where $r$ is the discount rate. $q$ is the price of capital. $c$ is the cost of capital services. $\delta$ is the depreciation rate. $t$ is the time of capital acquisition. $s$ is the time at which capital services are supplied. $u$ is the corporate tax rate. $k$ is the investment tax credit rate. And $D(s)$ is the proportion of the original cost of an asset of age $s$ that can be deducted from taxable income. Let $z$ be the present value of the depreciation deduction for one dollar of investment (after the tax credit),
\[
	z = \int_{0}^{\infty} e^{-rs}D(s)ds
\]
Under static expectations, and a substitution of $s = s - t$, the equation for the price of capital simplifies to
\begin{align*}
	q & = \int_{0}^{\infty} e^{-rs} ((1-u)c e^{-\delta s} + u(1- \theta k) q D(s))ds + kq \\
	  & = (1-u) c \int_{0}^{\infty} e^{-s(r + \delta)} ds + u(1- \theta k) q \int_{0}^{\infty} e^{-rs} D(s) ds + kq \\
	  & = \frac{(1 - u) c}{r + \delta} + u(1 - \theta k) q z + kq
\end{align*}
Simplifying this expression provides the following user cost of capital,
\[
	c = \underbrace{q (r + \delta)}_{\text{pre-tax rental price}} \cdot \underbrace{\frac{1 - k - u z + u \theta k z}{1-u}}_{\text{tax adjustment}}
\]
This equation is similar to the user cost of capital from \cite{hall_tax_1967}, the only difference is in the tax adjustment's numerator which includes $ - k +  u \theta k z$. Intuitively, the ITC tax rate reduces the cost of capital directly, $-k$. Simultaneously, the deduction of the ITC from the basis of the depreciation, $u \theta k z$, reduces the tax shield of investing, increasing the cost of capital. Note that the user cost of capital from \cite{hall_tax_1967} can be recovered with $\theta = 1$.

\subsubsection{Estimable ITC Elasticity}
To estimate the ITC elasticity, I am interested in the investment, $I$, responses with respect to the user cost of capital, $ucc$. 
\[
	\epsilon_{I} = \frac{\Delta \log(I)}{\Delta \log(ucc)}
\]
$\Delta \log(I)$ comes from estimates of Equation \ref{eq:did_poisson}. $\Delta \log(ucc)$ can be solved for using the derived user cost of capital and accounting for half of the ITC being deducted from the depreciation basis of investment, $\theta = 0.5$. Let the ITC rate in treatment counties be $k_{EC}$ and the ITC rate in control counties be $k_0$, then
\begin{align*}
	\Delta \log(ucc) & = \log\left(q (r + \delta) \cdot \frac{1 - k_{EC} - u z + u \theta k_{EC} z}{1-u}\right) - \log\left(q (r + \delta) \cdot \frac{1 - k_{0} - u z + u \theta k_{0} z}{1-u}\right) \\
	& = \log\left(\frac{1 - k_{EC} - u z + u \theta k_{EC} z}{1-u}\right) - \log\left( \frac{1 - k_{0} - u z + u \theta k_{0} z}{1-u}\right) \\
	& = \log\left(\frac{1 - k_{EC} - u z + u \theta k_{EC} z}{1 - k_{0} - u z + u \theta k_{0} z}\right)
\end{align*}
For estimation, I use the baseline ITC rate of $k_0 = 30\%$ in 2023 and $k_{EC} = 40\%$ with the 10 percentage point bonus from Energy Communities. For the corporate tax rate, I use the federal rate of $u =21\%$ in 2023. Since only half of the ITC had to be deducted from the depreciation basis of a project in 2023, I use $\theta = 0.5$. Finally, I need to create an estimate for $z$, the present value of depreciation deductions for \$1 of investment. 

Renewable energy projects face different depreciation schedules depending on their fuel source. For the ITC, I will calculate the present value of deductions for solar, the main recipient of ITC credits. In 2023, solar projects were subject to the Modified Accelerated Cost Recovery System (MACRS), 5-Year, Double Declining Balance depreciation schedule and bonus depreciation. The standard depreciation deduction schedule under MACRS, 5-Year, Double Declining Balance is presented in Column 2 of Table \ref{tab:macrs_5yr}. Bonus depreciation in 2023 allowed renewable energy investors to immediately deduct 80\% of a project's cost before applying the appropriate MACRS rate to the remaining 20\% of the cost basis. Column 3 of Table \ref{tab:macrs_5yr} presents the depreciation schedule for a \$1 million investment project without bonus depreciation, and Column 4 of Table \ref{tab:macrs_5yr} presents the depreciation schedule with bonus depreciation. Bonus depreciation significantly accelerates deductions to the first year an investment is placed in service allowing a project to deduct 84\% of an investment's cost in the first year compared to 20\% without bonus depreciation. Using the bonus depreciation rates, we can now estimate a discrete version of the present value of depreciation deduction for \$1 of investment.
\[
	z = \sum_{s=0}^{\infty} \frac{D(s)}{(1+r)^s}
\]
Under MACRS, 5-Year, Double Declining Balance with bonus depreciation and $r = 7\%$, $z = 0.91$. Then, under the Energy Community bonus the user cost of capital decreased by $18\%$. Readers can apply their own assumed tax changes to the raw estimates of investment responses. 

\begin{table}
\centering
\begin{tabular}{lccc}
	\toprule
	Year & \% of Depreciation Basis & \$1 mil. Basis & \$1 mil. Basis w/ Bonus Depreciation\\
	\midrule
	1	& 20.00\% & \$200,000 & \$840,000 \\
	2	& 32.00\% & \$320,000 & \$64,000 \\
	3	& 19.20\% & \$192,000 & \$38,400\\
	4	& 11.52\% & \$115,200 & \$23,040\\
	5	& 11.52\% & \$115,200 & \$23,040\\
	6	& 5.76\%  & \$57,600  & \$11,520\\
	\bottomrule
\end{tabular} 
\caption{\textbf{MACRS, 5-year, Double Declining Balance Depreciation Schedule.}}
\label{tab:macrs_5yr}
\end{table}

\end{document}